# Framework for an Intelligent Affect Aware Smart Home Environment for Elderly People

**Nirmalya Thakur**  *thakurna@mail.uc.edu*
*Department of Electrical Engineering and Computer Science*
*University of Cincinnati*
*Cincinnati, OH 45221-0030, USA*

**Chia Y. Han**  *han@ucmail.uc.edu*
*Department of Electrical Engineering and Computer Science*
*University of Cincinnati*
*Cincinnati, OH 45221-0030, USA*

**Abstract**

The population of elderly people has been increasing at a rapid rate over the last few decades and their population is expected to further increase in the upcoming future. Their increasing population is associated with their increasing needs due to problems like physical disabilities, cognitive issues, weakened memory and disorganized behavior, that elderly people face with increasing age. To reduce their financial burden on the world economy and to enhance their quality of life, it is essential to develop technology-based solutions that are adaptive, assistive and intelligent in nature. Intelligent Affect Aware Systems that can not only analyze but also predict the behavior of elderly people in the context of their day to day interactions with technology in an IoT-based environment, holds immense potential for serving as a long-term solution for improving the user experience of elderly in smart homes. This work therefore proposes the framework for an Intelligent Affect Aware environment for elderly people that can not only analyze the affective components of their interactions but also predict their likely user experience even before they start engaging in any activity in the given smart home environment. This forecasting of user experience would provide scope for enhancing the same, thereby increasing the assistive and adaptive nature of such intelligent systems. To uphold the efficacy of this proposed framework for improving the quality of life of elderly people in smart homes, it has been tested on three datasets and the results are presented and discussed.

## 1. INTRODUCTION

The ever-increasing population of elderly people has been one of the characteristics of this modern century. Currently there are around 962 million elderly people [1] across the world and they account for nearly 8.5 percent of the world's total population as shown in Figure 1. Recent studies [2] have predicted that by the year 2050 the population of elderly people will become around 1.6 billion globally and will end up outnumbering the population of younger people worldwide. Their number is further expected to increase and reach 3.1 billion by the year 2100.

Nirmalya Thakur & Chia Y. Han

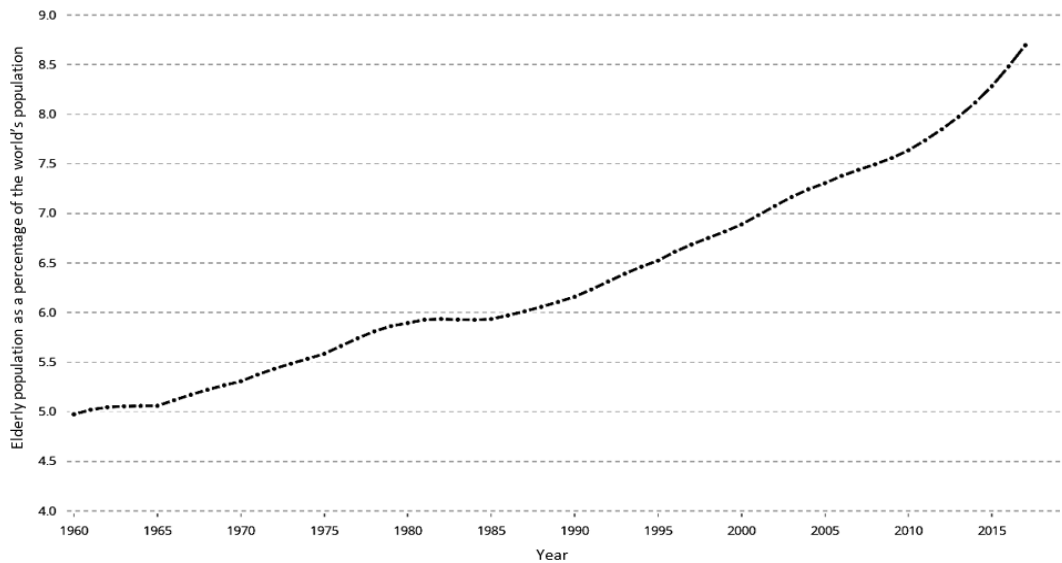

**FIGURE1:** The population of elderly people shown as a percentage of the global population, according to world population data by World Bank [3].

With increasing age, the various needs in terms of personal, social and healthcare requirements increase. The number of elderly people across the world with dementia has doubled in recent times [4] and their number is predicted to again double by the year 2030, leading to approximately 76 million people with dementia worldwide. In 2010 alone, approximately $604 billion costs were incurred to the healthcare industry in looking after people with dementia and this number is increasing at an alarming rate [4]. It is essential for modern day policy makers to develop systems and technologies that are equipped with functionalities to adapt and address these needs of elderly people. The challenges for policy makers in this context, is to design technologies that are affordable and sustainable and can help to improve the quality of life experienced by elderly people.

"The United Nations Principles for Older Persons" [5] adopted by United Nations General Assembly has identified a number of characteristics to determine and improve the quality of life experienced by elderly people. Some of the prominent ones are independence, participation, proper care, self-fulfillment and dignity. In addition to these, these principles also include the necessity to provide intelligent environments that are adaptable with respect to the dynamic needs of the elderly population and the need to develop technologies that can help elderly people retain their level of physical, mental and emotional well being in the context of their day to day routine activities.

According to recent reports of the World Health Organization, the shortage of caregivers is one of the issues in this regard. The current number of caregivers across the world is only 7.2 million and their population is expected to increase to around 12.9 million by 2035, which is significantly less as compared to the large scale predicted population increase of elderly people [6]. Smart home technology has been the focus of recent researchers to provide technology-based solutions to support elderly care and address their dynamic needs with increasing age. In an Internet of Things (IoT)-based smart home environment, the ability of smart devices to interact with the users and communicate with each other is being envisioned as a major resource to enhance the quality of life experienced by elderly people.

Recent researches [7-18] in the field of human computer interaction and affective computing for improving the quality of life experienced by elderly people have mostly focused on developing systems that can analyze multimodal aspects of user interactions to understand the underlining user experience. These technologies possess limited functionalities to enhance the user experience of the given activity or task being performed. Therefore, this paper proposes a framework for implementation of an intelligent affect aware smart home environment that can predict user experiences of activities even before users start engaging in those. This framework would provide means to enhance the interactions between users and systems and ensure adaptive and ambient living spaces for enhancing the quality of life experienced by elderly people in smart homes. This paper is organized as follows: Section II provides an overview of related works in this field. Section III provides details about the proposed framework which is followed by Section IV which discusses the results and findings. The conclusion and scope for future work is presented in Section V which is followed by references.

## 2. LITERATURE REVIEW

An indoor navigation system based on sensor networks, in the context of a smart home was proposed by Abascal et al. [7]. In this research, sensor technology was used to understand the instantaneous position of the user in the given environment and the same information was used by an intelligent wheelchair to help elderly people with disabilities navigate the given environment. This research also proposed the use of a handheld device that operated based on sensor data, to help elderly people with cognitive issues navigate properly in the given IoT space.

Chan et al. [8] proposed a model for monitoring the behavior of elderly in the context of their multimodal interactions with devices in a smart home. This system possessed the functionality to alert both formal and informal caregivers in the event of a fall. Yared et al. [9] developed an intelligent cooking environment to prevent any accidents to elderly people taking place in the kitchen. This work involved developing a knowledge base which consisted of the normal readings of various parameters of appliances present in the kitchen area. The system was equipped with the ability to understand any deviations from these normal running conditions to detect the condition of an accident in the kitchen.

Kim et al. [10] proposed an RFID based location tracking system to ensure the safety of elderly people in a smart home environment. The system had a knowledge base developed by monitoring people with respect to the average time they spend while interacting with different devices and systems in the context of their day to day living. The location tracking system was able to identify the location of the user in the given environment and analyze if the user was spending more time for performing a specific task with an aim to improve the mental well being of the user.

Deen [11] used a combination of sensors, wireless communication systems, electronic devices and intelligent computing technology to develop a system that can measure different physiological signals from the walking patterns of elderly people, with an aim to monitor their behavior and alert medical practitioners for any event that needed their attention. Civitarese et al. [12] proposed an intelligent system that could analyze multimodal aspects of user interactions to infer about elderly people successfully completing Activities of Daily Living (ADLs). The system was able to analyze user interactions in this context to understand about symptoms of Mild Cognitive Impairment (MCI) in elderly people.

Iglesias et al. [13] proposed a system that could collect the health information about elderly people by analyzing their interactions with touch-based interfaces. The system was also able to use wireless sensors to monitor them to analyze events of drastic changes in this health information to alert caregivers for addressing their needs and issues. Angelini et al. [14] proposed a smart bracelet to be used by elderly people while performing activities both in indoor and outdoor environments. This bracelet could collect information about the health status of the individual and alert the user of abnormal conditions. The device also possessed the ability to remind the user of medications and other daily routine tasks to enhance the overall well being of the user. Khosla et al. [15] conducted usability studies to propose a social robot that could

Nirmalya Thakur & Chia Y. Han

interact with elderly people in a smart home to help them perform different Activities of Daily Living (ADLs).

Tarik et al. [16] developed an intelligent activity recognition system for helping elderly people perform different activities in a smart home environment. The activity recognition system was equipped with the functionality to analyze a given activity being performed by the user and it also possessed the ability to perform intelligent decision making on whether to assist the user in the given activity being performed. Sarkar [17] proposed an intelligent robot called "NurseBot" that consisted of a scheduler to maintain the information about different medications that need to be taken by elderly people. It also possessed the functionality to deliver these specific medicines to elderly people as per their requirement. In a recent research by Thakur et al. [18], an intelligent system called CABERA – A Complex Activity Based Emotion Recognition Algorithm was proposed which could analyze the emotional response of the user in the context of different activities in a smart home to relate it to the underlining user experience.

Most of these works [7-14] have focused on multimodal ways of activity analysis with a focus on analyzing the user experience in the context of the given activity. Researchers [15, 17] have also focused on developing assistive robots to help users perform daily routine activities. This provides limited scope for enhancing the user experience of the given activity being performed by the user. The need to provide technology-based solutions that can not only analyze the user experience of different activities but also predict the same even before the activity is being performed, would serve as an essential characteristic of affect aware assistive technology to improve the quality of life experienced by elderly people in a smart home environment. This serves as the main motivation for this work.

This proposed framework extends the work done by Thakur et al. [18] which involved a methodology for forecasting user experiences in affect aware systems. The work in [18] was based on using probabilistic reasoning principles to develop rules between emotion, mood and outcome of any given activity to train a learning model to forecast the user experience. The main limitations of that work are (1) It is confined to a specific dataset and relevance of the work for other user interaction datasets or real time implementation is not discussed by the authors (2) The authors do not discuss the performance accuracy of the model that they use to infer emotions associated with different activities (3) The learning approach used by the authors does not yield a very high performance accuracy and (4) The authors did not discuss the relevance of the proposed learning method and whether or not it outperforms the other learning approaches. This proposed framework not only addresses these concerns but also discusses the approach for implementation of the proposed framework in any user interaction dataset as well as for real time user data.

A related work – Complex Activity Recognition Algorithm (CARALGO) has been used to propose this framework. Using CARALGO [19], any complex activity ($WCAtk$) can be broken down into small actions or tasks – these are called atomic activities ($At$) and the context parameters that affect these atomic activities are called context attributes ($Ct$). Each of these atomic activities and context variables are associated with specific weights based on probabilistic reasoning. Each complex activity has a set of specific atomic activities that are essential for performing the activity – these are called core atomic activities ($\gamma At$) and the context parameters affecting them are called core context attributes ($\rho Ct$). Based on the weights of atomic activities and context attributes associated to the complex activity, every complex activity is associated with a threshold function ($WTCAtk$) that helps to determine the occurrence of that activity. The total weight for any given occurrence of this complex activity should be equal to or greater than the value of its threshold function for the complex activity to have been successfully performed. In the event when the weight is less than the value of the threshold function, it helps to infer that the activity was not completed successfully by the user which could be due to several factors. CARALGO also helps to identify the start atomic activities ($AtS$), start context attributes ($CtS$), end atomic activities ($AtE$) and end context attributes ($CtE$) related to a complex activity.

## 3. PROPOSED WORK AND RESULTS
Implementation of this proposed framework for predicting user experiences in daily routine activities in an IoT-based smart home environment consists of the following steps:

1. Develop a database of user interactions in the context of day to day activities and ADLs in a smart and connected IoT-based environment
2. Identify specific characteristics of complex activities by analyzing their atomic activities and context attributes
3. Use CARALGO to infer about the start of any given complex activity
4. For each atomic activity performed on its associated context attribute
    a. Capture the users facial image on completion of the atomic activity
    b. Analyze the emotion conveyed by the facial image
    c. Record the information
5. Use CARALGO to infer about the end or completion of the given complex activity
6. Infer the overall emotion conveyed by the user by analyzing the above recorded information
7. Relate the emotion to the mood using specific set of rules obtained by a learning approach
8. Develop a learning approach to predict the user experience of activities based on this emotion and mood

To evaluate the efficacy of this proposed framework, it was implemented in RapidMiner [25]. RapidMiner is a data science software platform that provides an integrated development environment for implementing data science, machine learning, deep learning and natural language processing algorithms. The GUI of RapidMiner enables its users to develop these algorithms using the inbuilt "processes" in the desired manner. These "processes" can be developed by connecting the various "operators" that are provided by the software. An "operator" in RapidMiner is like a predefined method or function that consists of the general outline of a given task and can be modified by the user as per the requirement. A number of "operators" can be connected together to implement a specific functionality in the given "process". There are currently two versions of RapidMiner which are available – the free version and the commercial version. For implementation of this framework, the free version of RapidMiner was used.

Implementation of the above framework can be broadly summarized into three major subtasks once the dataset of user interactions consisting of ADLs in a smart home has been developed. The first subtask involves using CARALGO for activity recognition, identification of complex activities, determination of atomic activities and context attributes associated to all the given ADLs in the dataset. The second subtask involves analyzing the facial expressions for each of these atomic activities and identifying the emotional response as a whole, for the given complex activity. Thereafter the third task involves using a learning approach to relate the emotional response of the user to the mood to predict the user experience of any activity to be performed by the user next. These are discussed in the three sub sections that follow.

### 3.1 Using CARALGO to Analyze Complex Activities
For implementing CARALGO for activity recognition and activity analysis, a dataset was created which was a subset of the dataset developed by Sztyler et al. [20]. This dataset was developed during research on activity recognition at the University of Mannheim by Sztyler et al. [20]. The dataset consisted of tracking ADLs performed by seven individuals in an IoT-based smart home environment over a period of 14 days. The individuals were males and were aged around 23 years at the time of the experiment.

Multiple occurrences of the different ADLs performed by these individuals were captured using a host of wireless and wearable sensors as described in [20]. The specific subset of this dataset chosen for this work, consisted of details of one individual performing the different complex activities of grooming, deskwork, socializing, eating, drinking, going out for shopping, playing indoor sports and making meals. Multiple occurrences of these complex activities over a typical 24-hour period is shown in Figure 2.

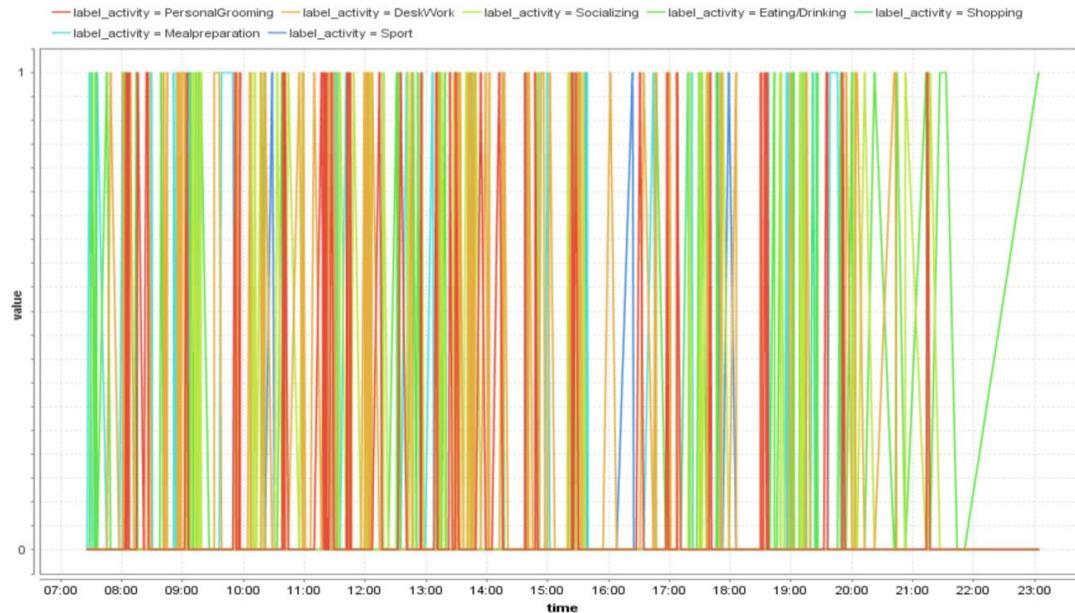

**FIGURE 2:** Analysis of multiple complex activities performed by one of the subjects in [20]. These different complex activities involve activities of grooming, deskwork, socializing, eating, drinking, going out for shopping, playing indoor sports and making meals.

Analysis by CARALGO for identification of all atomic activities and all context attributes, start atomic activities and start context attributes, end atomic activities and end context attributes, core atomic activities and core context attributes associated with these respective complex activities is shown in Tables 1-7. For each of these activities, the user is first assumed to be seated on a chair, after which the user initiates the given complex activity.

This analysis by CARALGO involved identifying all the small actions or tasks that the user performed towards reaching the end goal of completing the given activity. It also involved analyzing the context parameters on which these tasks were performed in the given IoT environment. These small actions or tasks are referred to as atomic activities and the context parameters on which these atomic activities were performed are called context attributes. Based on the relevance of these tasks and their associated context parameters, they were assigned weights by probabilistic reasoning principles.

For instance, Table 5 represents the CARALGO analysis of the complex activity of Going out for Shopping. The respective atomic activities are At1: Standing, At2: Putting on dress to go out, At3: Carrying bag, At4: Walking towards door, At5: Going out of door and their associated weights are At1: 0.12, At2: 0.32, At3: 0.30, At4: 0.13 and At5: 0.13. The context attributes associated with these respective atomic activities are Ct1: Lights on, Ct2: Dress Present, Ct3: Bag present, Ct4: Exit Door and Ct5: Door working and their associated weights are Ct1: 0.12, Ct2: 0.32, Ct3: 0.30, Ct4: 0.13 and Ct5: 0.13. As observed from this analysis the atomic activities At1 and At2 describe the user starting the complex activity and therefore they are identified as start atomic activities and their associated context attributes Ct1 and Ct2 are identified as start context attributes. Similarly, At4 and At5 describe the user completing the activity so they are identified as the end atomic activities and end context attributes respectively. Also, as the atomic activities At2 and At3 have the highest weights, they are identified as the core atomic activities and their associated context parameters are identified as core context attributes for this complex activity.

Nirmalya Thakur & Chia Y. Han**TABLE 1:** Analysis by CARALGO of the Complex Activity of Grooming (AG).

| | |
|---|---|
| Complex Activity WCAtk (WT Atk) - AG (0.67) | |
| Weight of Atomic Activities WtAti | At1: Standing (0.14) At2: Walking towards Mirror (0.16) At3: Picking up grooming kit (0.25) At4: Taking out grooming instrument (0.15) At5: Sitting Down to groom (0.08) At6: Putting back grooming kit (0.12) At7: Coming back to seat (0.10) |
| Weight of Context Attributes WtCti | Ct1: Lights on (0.14) Ct2: Mirror present (0.16) Ct3: Presence of grooming kit (0.25) Ct4: Grooming instruments present (0.15) Ct5: Sitting Area (0.08) Ct6: Presence of space for kit (0.12) Ct7: Seating area (0.10) |
| Core γAt and ρCt | At2,At3,At4 and Ct2,Ct3,Ct4 |
| Start AtS and CtS | At1, At2 and Ct1, Ct2 |
| End AtE and CtE | At5, At6, At7 and Ct5, Ct6, Ct7 |

**TABLE 2:** Analysis by CARALGO of the Complex Activity of Working on a Desk (WD).

| | |
|---|---|
| Complex Activity WCAtk (WT Atk) - UL (0.82) | |
| Weight of Atomic Activities WtAti | At1: Standing (0.10) At2: Walking Towards Desk (0.23) At3: Taking out Laptop (0.28) At4: Typing log in password (0.15) At5: Sitting Down near Laptop (0.06) At6: Opening Required Application (0.10) At7: Connecting any peripheral devices like mouse, keyboard etc. (0.08) |
| Weight of Context Attributes WtCti | Ct1: Lights on (0.10) Ct2: Desk Area (0.23) Ct3: Laptop Present (0.28) Ct4: Log-in feature working (0.15) Ct5: Sitting Area (0.06) Ct6: Required Application Present (0.10) Ct7: Peripheral devices (0.08) |
| Core γAt and ρCt | At2, At3 and Ct2, Ct3 |
| Start AtS and CtS | At1, At2, and Ct1, Ct2 |
| End AtE and CtE | At6, At7 and Ct6, Ct7 |

**TABLE 3:** Analysis by CARALGO of the Complex Activity of Socializing with Friends (SF).

| | |
|---|---|
| Complex Activity WCAtk (WT Atk) - LTS (0.70) | |
| Weight of Atomic Activities WtAti | At1: Standing (0.16) At2: Walking Towards Main Door (0.18) At3: Welcoming Friends (0.23) At4: Going to Entertainment Area (0.13) At5: Seating Down (0.11) At6: Starting a discussion (0.11) At7: Serving food (0.08) |
| Weight of Context Attributes WtCti | Ct1: Lights on (0.16) Ct2: Main Door Area (0.18), Ct3: Friends Present (0.23) Ct4: Entertainment Area (0.13) Ct5: Seating Space Available (0.11) Ct6: Discussion Topic (0.11) Ct7: Food Present (0.08) |
| Core γAt and ρCt | At2, At3, At4 and Ct2, Ct3, Ct4 |
| Start AtS and CtS | At1, At2 and Ct1, Ct2 |
| End AtE and CtE | At6, At7 and Ct6, Ct7 |

**TABLE 4:** Analysis by CARALGO of the Complex Activity of Enjoying Drinks (ED).

| Complex Activity WCAtk (WT Atk) – EL (0.72) | |
|---|---|
| Weight of Atomic Activities WtAti | At1: Standing (0.08) At2: Walking towards drinking area (0.20) At3: Taking out the drink to be poured (0.25) At4: Preparing the drink (0.20) At5: Sitting down (0.08) At6: Starting to drink (0.19) |
| Weight of Context Attributes WtCti | Ct1: Lights on (0.08) Ct2: Drinking Area (0.20) Ct3: Drink present (0.25) Ct4: Drinking cup present (0.20) Ct5: Sitting options available (0.08) Ct6: Drink taste (0.19) |
| Core γAt and ρCt | At2, At3, At4 and Ct2, Ct3, Ct4 |
| Start AtS and CtS | At1, At2 and Ct1, Ct2 |
| End AtE and CtE | At5, At6 and Ct5, Ct6 |

**TABLE 5:** Analysis by CARALGO of the Complex Activity of Going out for Shopping (GS).

| Complex Activity WCAtk (WT Atk) - GS (0.68) | |
|---|---|
| Weight of Atomic Activities WtAti | At1: Standing (0.12) At2: Putting on dress to go out (0.32) At3: Carrying bag (0.30) At4: Walking towards door (0.13) At5: Going out of door (0.13) |
| Weight of Context Attributes WtCti | Ct1: Lights on (0.12), Ct2: Dress Present (0.32), Ct3: Bag present (0.30), Ct4: Exit Door (0.13), Ct5: Door working (0.13) |
| Core γAt and ρCt | At2, At3 and Ct2, Ct3 |
| Start AtS and CtS | At1, At2 and Ct1, Ct2 |
| End AtE and CtE | At4, At5 and Ct4, Ct5 |

**TABLE 6:** Analysis by CARALGO of the Complex Activity of Playing Indoor Games (PIG).

| Complex Activity WCAtk (WT Atk) - LV (0.68) | |
|---|---|
| Weight of Atomic Activities WtAti | At1: Standing (0.12) At2: Going to playing area (0.32) At3: Taking out playing equipment's (0.30) At4: Inviting others to play (0.13) At5: Starting to play (0.13) |
| Weight of Context Attributes WtCti | Ct1: Lights on (0.12), Ct2: Playing Area (0.32), Ct3: Playing equipment's present (0.30), Ct4: Other players present (0.13), Ct5: Game knowledge (0.13) |
| Core γAt and ρCt | At2, At3 and Ct2, Ct3 |
| Start AtS and CtS | At1, At2 and Ct1, Ct2 |
| End AtE and CtE | At4, At5 and Ct4, Ct5 |

**TABLE 7:** Analysis by CARALGO of the Complex Activity of Making Meals (MM).

| Complex Activity WCAtk (WT Atk) - MB (0.73) | |
|---|---|
| Weight Of Atomic Activities WtAti | At1: Standing (0.10) At2: Walking Towards Kitchen (0.12) At3: Loading Food In Microwave (0.14) At4: Turning on Microwave (0.25) At5: Setting The Time (0.15) At6: Taking out prepared meal (0.18) At7: Sitting down to eat (0.06) |
| Weight Of Context Attributes WtCti | Ct1: Lights on (0.10), Ct2: Kitchen Area (0.12), Ct3: Microwavable food Present (0.14), Ct4: Time settings (0.15), Ct5: Microwave Present (0.25), Ct6: Microwave Working (0.18), Ct7: Sitting down to eat (0.06) |
| Core γAt and ρCt | At4, At5, At6 and Ct4, Ct5, Ct6 |
| Start AtS and CtS | At1, At2 and Ct1, Ct2 |
| End AtE and CtE | At6, At7 and Ct6, Ct7 |

Nirmalya Thakur & Chia Y. Han

**3.2 Facial Emotion Analysis for Atomic Activities**
For implementation of this framework to analyze the emotional response of each atomic activity and its associated context attribute for a given complex activity, the following steps need to be implemented:
1. Capture the facial image of the user after completion of each atomic activity.
2. Represent the facial image by a fixed number of feature points and record muscle movements at each of these feature points.
3. Develop a database of these images and use methodology of Facial Action Coding System (FACS) to associate each of them with one of the seven basic emotional states Angry, Disgust, Fear, Happy, Sad, Surprise, Neutral.
4. Train a learning model based on this data to enable it to classify new facial images into the various emotional states.
5. Capture the facial images during a given complex activity by identifying the start and end of the complex activity with CARALGO.
6. Use this learning model to analyze these facial images and evaluate the overall emotional state during the complex activity occurence.
7. Relate the emotional state to the mood using a specific set of rules.
8. Develop a learning model that can predict user experiences based on mood, emotional response and the outcome of the given activity.

To evaluate the efficacy of the framework in implementing the above functionality, a dataset was developed based on the dataset which was a result of the work of Goodfellow et al. in [21]. This work by Goodfellow et al. [21] consists of facial images of users captured while they performed different activities. The respective images in this dataset also provide information about the muscle movements and outline the associated emotional state using Facial Action Coding System (FACS). Using this information, a learning model was developed in RapidMiner that can analyze this information and classify images into the seven basic emotional states - Angry, Disgust, Fear, Happy, Sad, Surprise, Neutral. The system first consisted of using operators in RapidMiner to preprocess the data, detect and remove outliers, spilt the data into training set and test set, which was followed by training the model. 80% of the data was used for training the model and 20% of the data was used as the test data. This system is shown in Figure 3.

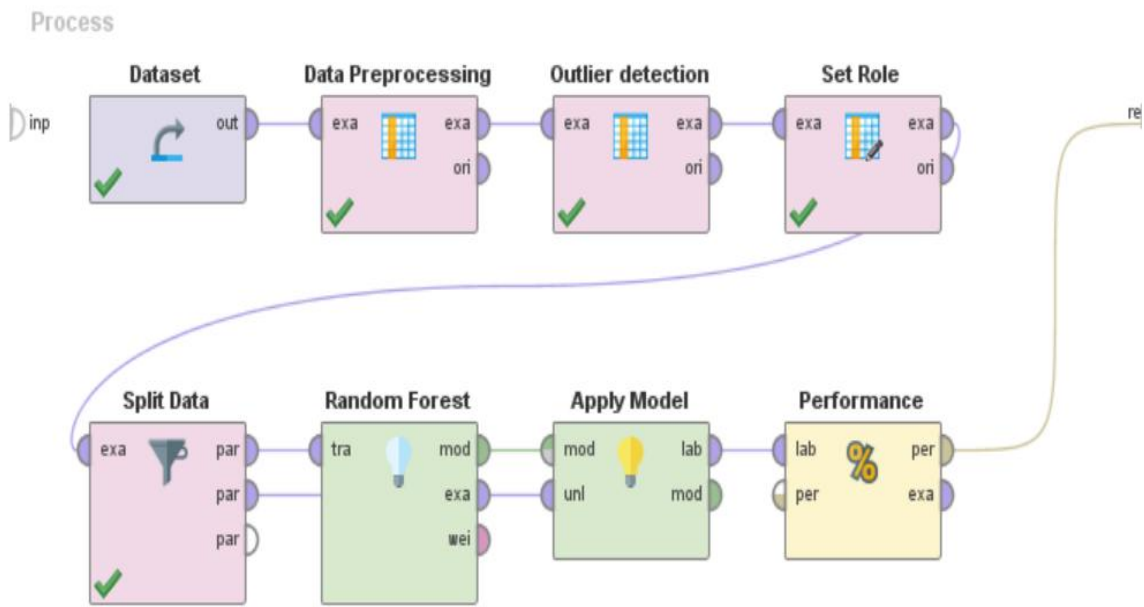

**FIGURE 3:** System developed in RapidMiner for analyzing emotions from facial images.

Different learning approaches were built in RapidMiner and their respective performance accuracies were compared to choose the learning approach that had the highest performance accuracy. Some of the major learning approaches – neural networks, decision tree learning, random forest, naïve bayes, deep learning and K-NN classifier were implemented as shown in Table 8. The seven different emotional states were considered as seven classes and the highest-class precision, lowest class precision and overall performance accuracies of these learning methods for this chosen dataset were compared.

These performance characteristics were obtained by analyzing the working of these learning methods through confusion matrices. Table 8 shows the comparison of the performance characteristics of these learning methods, details about their respective performance accuracies are presented in Figures 4-9 in the form of confusion matrices.

**TABLE 8:** Comparison of Different Learning Methods for Analyzing Emotions from Facial Images from this Dataset.

| Learning Method Name | Overall Performance Accuracy | Lowest Class Precision | Highest Class Precision |
|---|---|---|---|
| Artificial Neural Networks | 34.69% | 29.28% | 42.12% |
| Decision Tree Learning | 77.50% | 33.03% | 100.00% |
| Random Forest | 79.53% | 49.55% | 100.00% |
| Naïve Bayes | 69.30% | 47.29% | 100.00% |
| Deep Learning | 38.78% | 11.13% | 50.06% |
| K-NN Classifier | 69.37% | 46.63% | 83.30% |

accuracy: 34.69%

| | true angry | true disgust | true sad | true neutral | true happy | true fear | true surprise | class precision |
|---|---|---|---|---|---|---|---|---|
| pred. angry | 0 | 0 | 0 | 0 | 0 | 0 | 0 | 0.00% |
| pred. disgust | 0 | 0 | 0 | 0 | 0 | 0 | 0 | 0.00% |
| pred. sad | 0 | 0 | 0 | 0 | 0 | 0 | 0 | 0.00% |
| pred. neutral | 0 | 0 | 0 | 0 | 0 | 0 | 0 | 0.00% |
| pred. happy | 0 | 112 | 893 | 112 | 893 | 0 | 110 | 42.12% |
| pred. fear | 1117 | 0 | 0 | 670 | 112 | 1005 | 448 | 29.98% |
| pred. surprise | 0 | 0 | 0 | 0 | 0 | 0 | 0 | 0.00% |
| class recall | 0.00% | 0.00% | 0.00% | 0.00% | 88.86% | 100.00% | 0.00% | |

**FIGURE 4:** Performance Characteristics in the form of a confusion matrix for the learning system (Artificial Neural Networks) developed for analyzing emotions from facial images from this dataset.

accuracy: 77.50%

| | true angry | true disgust | true sad | true neutral | true happy | true fear | true surprise | class precision |
|---|---|---|---|---|---|---|---|---|
| pred. angry | 1005 | 0 | 0 | 0 | 0 | 0 | 112 | 89.97% |
| pred. disgust | 0 | 0 | 0 | 0 | 0 | 0 | 0 | 0.00% |
| pred. sad | 0 | 0 | 669 | 0 | 0 | 0 | 0 | 100.00% |
| pred. neutral | 0 | 112 | 112 | 782 | 112 | 0 | 0 | 69.95% |
| pred. happy | 0 | 0 | 112 | 0 | 670 | 0 | 0 | 85.68% |
| pred. fear | 112 | 0 | 0 | 0 | 0 | 1005 | 336 | 69.17% |
| pred. surprise | 0 | 0 | 0 | 0 | 223 | 0 | 110 | 33.03% |
| class recall | 89.97% | 0.00% | 74.92% | 100.00% | 66.67% | 100.00% | 19.71% | |

**FIGURE 5:** Performance Characteristics in the form of a confusion matrix for the learning system (Decision Tree Learning) developed for analyzing emotions from facial images from this dataset.

**accuracy: 79.53%**

|  | true angry | true disgust | true sad | true neutral | true happy | true fear | true surprise | class precision |
|---|---|---|---|---|---|---|---|---|
| pred. angry | 1005 | 0 | 0 | 0 | 0 | 0 | 112 | 89.97% |
| pred. disgust | 0 | 0 | 0 | 0 | 0 | 0 | 0 | 0.00% |
| pred. sad | 0 | 0 | 669 | 0 | 0 | 0 | 0 | 100.00% |
| pred. neutral | 0 | 112 | 112 | 782 | 112 | 0 | 0 | 69.95% |
| pred. happy | 0 | 0 | 112 | 0 | 781 | 0 | 0 | 87.46% |
| pred. fear | 112 | 0 | 0 | 0 | 0 | 1005 | 336 | 69.17% |
| pred. surprise | 0 | 0 | 0 | 0 | 112 | 0 | 110 | 49.55% |
| class recall | 89.97% | 0.00% | 74.92% | 100.00% | 77.71% | 100.00% | 19.71% |  |

**FIGURE 6:** Performance Characteristics in the form of a confusion matrix for the learning system (Random Forest) developed for analyzing emotions from facial images from this dataset.

**accuracy: 69.30%**

|  | true angry | true disgust | true sad | true neutral | true happy | true fear | true surprise | class precision |
|---|---|---|---|---|---|---|---|---|
| pred. angry | 333 | 0 | 0 | 0 | 0 | 0 | 112 | 74.83% |
| pred. disgust | 0 | 0 | 0 | 0 | 0 | 0 | 0 | 0.00% |
| pred. sad | 0 | 0 | 669 | 0 | 0 | 0 | 0 | 100.00% |
| pred. neutral | 0 | 112 | 112 | 782 | 112 | 0 | 0 | 69.95% |
| pred. happy | 0 | 0 | 112 | 0 | 893 | 0 | 0 | 88.86% |
| pred. fear | 784 | 0 | 0 | 0 | 0 | 1005 | 336 | 47.29% |
| pred. surprise | 0 | 0 | 0 | 0 | 0 | 0 | 110 | 100.00% |
| class recall | 29.81% | 0.00% | 74.92% | 100.00% | 88.86% | 100.00% | 19.71% |  |

**FIGURE 7:** Performance Characteristics in the form of a confusion matrix for the learning system (Naïve Bayes) developed for analyzing emotions from facial images from this dataset.

**accuracy: 38.78%**

|  | true angry | true disgust | true sad | true neutral | true happy | true fear | true surprise | class precision |
|---|---|---|---|---|---|---|---|---|
| pred. angry | 1117 | 0 | 0 | 223 | 0 | 1005 | 448 | 39.99% |
| pred. disgust | 0 | 0 | 0 | 0 | 0 | 0 | 0 | 0.00% |
| pred. sad | 0 | 112 | 112 | 559 | 112 | 0 | 0 | 12.51% |
| pred. neutral | 0 | 0 | 0 | 0 | 0 | 0 | 0 | 0.00% |
| pred. happy | 0 | 0 | 781 | 0 | 893 | 0 | 110 | 50.06% |
| pred. fear | 0 | 0 | 0 | 0 | 0 | 0 | 0 | 0.00% |
| pred. surprise | 0 | 0 | 0 | 0 | 0 | 0 | 0 | 0.00% |
| class recall | 100.00% | 0.00% | 12.54% | 0.00% | 88.86% | 0.00% | 0.00% |  |

**FIGURE 8:** Performance Characteristics in the form of a confusion matrix for the learning system (Deep Learning) developed for analyzing emotions from facial images from this dataset.

Nirmalya Thakur & Chia Y. Han

accuracy: 69.37%

| | true angry | true disgust | true sad | true neutral | true happy | true fear | true surprise | class precision |
|---|---|---|---|---|---|---|---|---|
| pred. angry | 1117 | 0 | 0 | 112 | 0 | 0 | 112 | 83.30% |
| pred. disgust | 0 | 0 | 0 | 0 | 0 | 0 | 0 | 0.00% |
| pred. sad | 0 | 112 | 781 | 670 | 112 | 0 | 0 | 46.63% |
| pred. neutral | 0 | 0 | 0 | 0 | 0 | 0 | 0 | 0.00% |
| pred. happy | 0 | 0 | 112 | 0 | 893 | 0 | 110 | 80.09% |
| pred. fear | 0 | 0 | 0 | 0 | 0 | 1005 | 336 | 74.94% |
| pred. surprise | 0 | 0 | 0 | 0 | 0 | 0 | 0 | 0.00% |
| class recall | 100.00% | 0.00% | 87.46% | 0.00% | 88.86% | 100.00% | 0.00% | |

**FIGURE 9:** Performance Characteristics in the form of a confusion matrix for the learning system (K NN Classifier) developed for analyzing emotions from facial images from this dataset.

As observed from Table 8 and from Figures 4-9, the Random Forest learning approach has the highest-class precision as well as the highest overall performance accuracy for this given dataset. Therefore, it was chosen to associate facial expressions with emotional states for development of this framework. The system implementing this Random Forest learning approach is shown in Figure 3 and its performance characteristics are shown in Figure 6.

**3.3 Predicting User Experience of Complex Activities**
This learning approach can therefore be used to classify the emotional response associated to different atomic activities and evaluate the overall emotional response of the complex activity. The overall emotional response of a given complex activity is inferred based on the overall emotional state associated with the given complex activity as per the following algorithm based on CARALGO:

while (WCAtk >= WTAtk)
{
while (AtS!=AtE)
{
*Capture Facial Image*
*Analyze Emotional Response = E*
if (E == Angry) -> nAngry++
else if (E == Disgust) ->nDisgust++
else if (E == Fear) -> nFear++
else if (E== Happy) -> nHappy++
else if (E== Sad) -> nSad++
else if (E== Surprise) -> nSurpise++
else if (E== Neutral) ->nNeutral++
}
If ((nAngry> nDisgust) && (nAngry> nFear) && (nAngry> nHappy)
   && (nAngry>nSad) && (nAngry> nSurprise) && (nAngry> nNeutral))
        {OE = Angry}
else if ((nDisgust>nAngry) && (nDisgust > nFear) && (nDisgust > nHappy)
        && (nDisgust>nSad) && (nDisgust> nSurprise) && (nDisgust> nNeutral))
         {OE = Disgust}
else if ((nFear>nAngry) && (nFear>nDisgust) && (nFear> nHappy)
        && (nFear>nSad) && (nFear> nSurprise) && (nFear> nNeutral))
         {OE = Fear}
else if ((nHappy>nAngry) && (nHappy>nDisgust) && (nHappy>nFear)
        && (nHappy>nSad) && (nHappy> nSurprise) && (nHappy> nNeutral))
         {OE = Happy}
else if ((nSad>nAngry) && (nSad>nDisgust) && (nSad>nFear)

```
        && (nSad>nHappy) && (nSad> nSurprise) && (nSad> nNeutral))
           {OE = Sad}
else if ((nSurprise>nAngry) && (nSurprise>nDisgust) && (nSurprise>nFear)
        && (nSurprise>nHappy) && (nSurprise>nSad) && (nSurprise> nNeutral))
           {OE = Surprise}
else if ((nNeutral>nAngry) && (nNeutral>nDisgust) && (nNeutral>nFear)
        && (nNeutral>nHappy) && (nNeutral>nSad) && (nNeutral>nSurprise))
           {OE = Neutral}
}
```

where
WCAtk = Weight of Complex Activity
WTAtk = Threshold Weight of Complex Activity
AtS = Start Atomic Activity
CtS = Start Context Attribute
AtE = End Atomic Activity
CtE = End Context Attribute
E = Emotional Response of the given atomic activity
nAngry = Number of atomic activities showing angry emotion
nDisgust = Number of atomic activities showing disgust emotion
nFear = Number of atomic activities showing fear emotion
nNeutral = Number of atomic activities showing neutral emotion
nHappy = Number of atomic activities showing happy emotion
nSurprise = Number of atomic activities showing surprise emotion
nSad = Number of atomic activities showing sad emotion
OE = Overall emotional response of the Complex Activity

This algorithm based on CARALGO initially checks for the condition for the occurrence of the complex activity by checking whether the weight associated to the complex activity exceeds its threshold weight. Thereafter, for each atomic activity, the emotion associated with the same is analyzed. Seven counter variables are used to store the number of times each of these seven respective emotions are recorded from facial images during the course of the entire complex activity. This information is thereafter used to identify that counter variable which has the greatest value as compared to all the other counter variables. The emotion associated to this counter variable is considered as the overall emotion associated to the complex activity.

The emotional response associated to the overall activity and the subsequent emotional responses as a result of performing different complex activities can be analyzed to evaluate the overall mood of the person as discussed in [18]. This involves analyzing the respective complex activities through CARALGO, evaluating their emotional response and then analyzing the emotional response of each activity to infer the overall mood of the user. A set of specific complex activities were analyzed from the UK Domestic Appliance Level Electricity (DALE) dataset [22] for implementation of this framework. The UK DALE data set consists of details of appliance usage patterns recorded over a time resolution of six seconds in five smart homes in Southern England over a span of three years from 2012-2015. A related work [23] is referred to relate these appliance usage patterns with complex activities. A specific set of complex activities from a typical day from this dataset, as shown in Table 9 were analyzed to validate the working of this framework. These activities included the complex activities of Cooking in Kitchen, Using Washing Machine, Doing office work, Watching TV and Making Breakfast which were performed on this typical day in between 10:00 AM and 12:30PM. The CARALGO [19] analysis of these complex activities is shown in Tables 10-14 along with their respective emotional responses [24], which is summarized in Table 15.

Nirmalya Thakur & Chia Y. Han

**TABLE 9:** A Typical Set of Activities between 10:00 AM to 12:30PM on a typical day obtained from the UK DALE Dataset for this Analysis.

| Time of day | Complex Activity Details |
|---|---|
| 10:00 AM to 10:30 AM | Cooking in Kitchen |
| 10:30 AM to 11:00 AM | Using Washing Machine |
| 11:00 AM to 11:30 AM | Doing office work |
| 11:30 AM to 12:00 PM | Watching TV |
| 12:00 PM to 12:30 PM | Making Breakfast |

**TABLE 10:** Analysis by CARALGO of the Complex Activity of Cooking in Kitchen (CFWK).

| Complex Activity WCAtk (WT Atk) - CFWK (0.60) | |
|---|---|
| Weight of Atomic Activities WtAti | At1: Standing (0.10) At2: Walking Towards Kitchen Area (0.16) At3: Loading Food Into Container (0.19) At4: Turning On Burner (0.12) At5: Adjusting Heat (0.09) At6: Adding spices in Food (0.11) At7: Stirring (0.07) At8: Turning Off burner (0.12) At9: Sitting Back (0.04) |
| Weight of Context Attributes WtCti | Ct1: Lights on (0.10) Ct2: Kitchen Area (0.16) Ct3: Food to be cooked (0.19) Ct4: Burner Turning On (0.12) Ct5: Heat Settings (0.09) Ct6: Food spices (0.11) Ct7: Stirrer (0.07) Ct8: Burner Turning off (0.12) Ct9: Sitting Area (0.04) |
| Core γAt and ρCt | At2, At3 and Ct2, Ct3 |
| Start AtS and CtS | At1, At2 and Ct1, Ct2 |
| End AtE and CtE | At8, At9 and Ct8, Ct9 |
| Emotional Response | Negative |

**TABLE 11:** Analysis by CARALGO of the Complex Activity of Using Washing Machine (UWM).

| Complex Activity WCAtk (WT Atk) - UWM (0.72) | |
|---|---|
| Weight of Atomic Activities WtAti | At1: Standing (0.08) At2: Walking Towards Machine (0.20) At3: Turning On Machine (0.25) At4: Pouring Detergent (0.08) At5: Loading Clothes (0.20) At6: Adjusting Timer (0.12) At7: Sitting Down (0.07) |
| Weight of Context Attributes WtCti | Ct1: Lights on (0.08) Ct2: Laundry Area (0.20) Ct3: Washing Machine present (0.25) Ct4: Detergent Available (0.08) Ct5: Presence of clothes (0.20) Ct6: Timer settings working (0.12) Ct7: Sitting Area (0.07) |
| Core γAt and ρCt | At2, At3, At5 and Ct2, Ct3, Ct5 |
| Start AtS and CtS | At1, At2 and Ct1, Ct2 |
| End AtE and CtE | At6, At7 and Ct6, Ct7 |
| Emotional Response | Negative |

Nirmalya Thakur & Chia Y. Han

**TABLE 12:** Analysis by CARALGO of the Complex Activity of Doing Office Work (DOW).

| Complex Activity WCAtk (WT Atk) - UL (0.82) | |
|---|---|
| Weight of Atomic Activities WtAti | At1: Standing (0.10) At2: Walking Towards Office Desk (0.15) At3: Turning on Laptop (0.28) At4: Typing log in password (0.23) At5: Sitting Down near Laptop (0.06) At6: Opening Required Application (0.10) At7: Connecting any peripheral devices like mouse, keyboard etc. (0.08) |
| Weight of Context Attributes WtCti | Ct1: Lights on (0.10) Ct2: Office Desk Area (0.15) Ct3: Laptop Present (0.28) Ct4: Log-in feature working (0.23) Ct5: Sitting Area (0.06) Ct6: Required Application Present (0.10) Ct7: Peripheral devices (0.08) |
| Core γAt and ρCt | At2, At3, At4 and Ct2, Ct3, Ct4 |
| Start AtS and CtS | At1, At2, and Ct1, Ct2 |
| End AtE and CtE | At6, At7 and Ct6, Ct7 |
| Emotional Response | Negative |

**TABLE 13:** Analysis by CARALGO of the Complex Activity of Watching TV (WT).

| Complex Activity WCAtk (WT Atk) - WT (0.67) | |
|---|---|
| Weight of Atomic Activities WtAti | At1: Standing (0.15) At2: Walking towards TV (0.15) At3: Turning on the TV (0.25) At4: Fetching the remote control (0.15) At5: Sitting Down (0.08) At6: Tuning Proper Channel (0.12) At7: Adjusting Display and Audio (0.10) |
| Weight of Context Attributes WtCti | Ct1: Lights on (0.15) Ct2: Entertainment Area (0.15) Ct3: Presence of TV (0.25) Ct4: Remote Control Available (0.15) Ct5: Sitting Area (0.08) Ct6: Channel Present (0.12) Ct7: Settings working (0.10) |
| Core γAt and ρCt | At2,At3,At4 and Ct2,Ct3,Ct4 |
| Start AtS and CtS | At1, At2 and Ct1, Ct2 |
| End AtE and CtE | At5, At6, At7 and Ct5, Ct6, Ct7 |
| Emotional Response | Positive |

**TABLE 14:** Analysis by CARALGO of the Complex Activity of Making Breakfast (MB).

| Complex Activity WCAtk (WT Atk) - MBUT (0.73) | |
|---|---|
| Weight of Atomic Activities WtAti | At1: Standing (0.10) At2: Walking Towards Toaster (0.12) At3: Putting bread into Toaster (0.15) At4: Setting the Time (0.15) At5: Turning off toaster (0.25) At6: Taking out bread (0.18) At7: Sitting Back (0.05) |
| Weight of Context Attributes WtCti | Ct1: Lights on (0.10), Ct2: Kitchen Area (0.12), Ct3: Bread Present (0.15), Ct4: Time settings working (0.15), Ct5: Toaster Present (0.25), Ct6: Bread cool (0.18), Ct7: Sitting Area (0.05) |
| Core γAt and ρCt | At3, At4, At5 and Ct3, Ct4, Ct5 |
| Start AtS and CtS | At1, At2 and Ct1, Ct2 |
| End AtE and CtE | At6, At7 and Ct6, Ct7 |
| Emotional Response | Negative |

**TABLE 15:** Comparison of Emotional Response of Different Complex Activities Performed Between 10AM TO 12:30PM on a Typical Day from the UK DALE dataset considered here.

| Time of day | Complex Activity Details | Emotional Response |
|---|---|---|
| 10:00 AM to 10:30 AM | Cooking in Kitchen | Negative |
| 10:30 AM to 11:00 AM | Using Washing Machine | Negative |
| 11:00 AM to 11:30 AM | Doing office work | Negative |
| 11:30 AM to 12:00 PM | Watching TV | Positive |
| 12:00 PM to 12:30 PM | Making Breakfast | Negative |

According to [18] as the mood of the user is the aggregate of the emotional responses shown by the user while performing some of the most recent complex activities, it can be concluded that as a result of this specific sequence of complex activities with their respective emotional responses, the overall mood of the user starting the next complex activity would be negative. This relationship between emotion and mood is very similar to the analogy between weather and climate. The intensity of these respective emotions also plays a significant role in having an impact on the overall mood and this relationship can be determined by assigning weights with the respective emotions. These weights would specify the intensity of the given emotion and analysis of the same would help in determining the overall mood of the user as well the intensity of the same.

Thereafter, the general rules [18] for relating the mood with the outcome of a given activity to predict the user experience of the same are used to develop a learning model. The outcome in this context refers to the condition of whether the user was able to complete the given activity, or the user failed to do so. Analysis of the same is done through CARALGO [19]. These rules are presented in Table 15.

**TABLE 16:** Relationship Between Mood and User Experience in Terms of Activity Outcome.

| Mood | Outcome | User Experience |
|---|---|---|
| Positive | Positive | Positive |
| Positive | Negative | Negative |
| Negative | Positive | Positive |
| Negative | Negative | Negative |

A number of complex activities from the UK DALE dataset [22] were analyzed using this framework to obtain their associated emotional responses in different contexts. Then the above rules as mentioned in Table 16 were used to train a learning model that could predict the user experience of these activities based on the users mood and outcome of the given activity. To evaluate the actual user experience of these respective complex activities, CABERA – A Complex Activity Based Emotion Recognition Algorithm proposed by Thakur et al. [24] was used. These results were compared to analyze the performance accuracy of the system. As per Table 8, Random Forest learning approach was chosen to develop this predictive model. The system as developed in RapidMiner is shown in Figure 10.

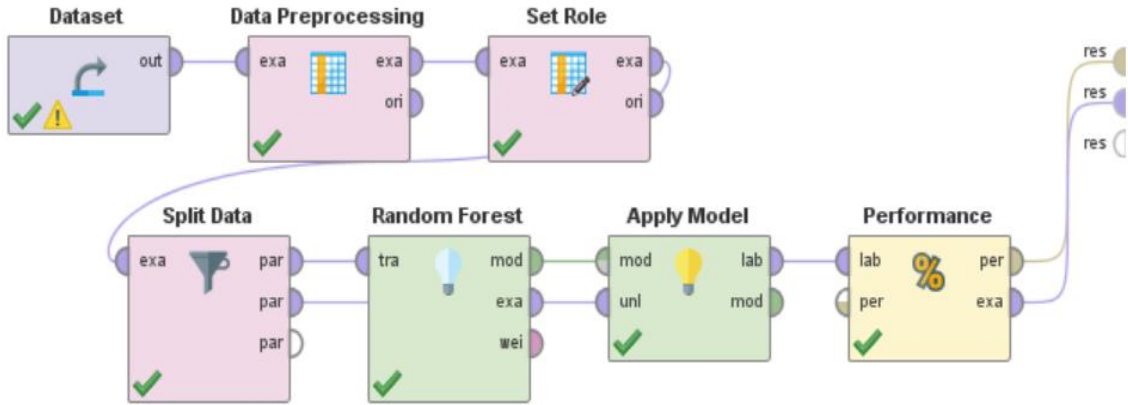

**FIGURE 10:** System developed in RapidMiner for predicting user experience based on nature of user interactions in complex activities.

The output of this system displayed the confidence value related to each prediction of the User Experience based on the values of mood and outcome. For simplicity of implementation just positive and negative values for each of the attributes – Mood, Outcome and User Experience, were used to develop this predictive model. The highest confidence valued attribute was used to make the prediction for a given condition to predict the User Experience. A screenshot of the output of this system showing these different attributes is shown in Figure 11. Figure 12 shows the performance characteristics of this system in the form of a confusion matrix.

| Row No. | UX | prediction(UX) | confidence(Positive) | confidence(Negative) | Mood | Outcome |
|---|---|---|---|---|---|---|
| 1 | Positive | Positive | 0.620 | 0.380 | Negative | Positive |
| 2 | Positive | Positive | 0.727 | 0.273 | Positive | Positive |
| 3 | Positive | Positive | 0.727 | 0.273 | Positive | Positive |
| 4 | Negative | Negative | 0.183 | 0.817 | Positive | Negative |
| 5 | Positive | Negative | 0.356 | 0.644 | Negative | Negative |
| 6 | Positive | Positive | 0.727 | 0.273 | Positive | Positive |
| 7 | Negative | Positive | 0.727 | 0.273 | Positive | Positive |
| 8 | Negative | Negative | 0.356 | 0.644 | Negative | Negative |
| 9 | Negative | Negative | 0.183 | 0.817 | Positive | Negative |
| 10 | Negative | Positive | 0.620 | 0.380 | Negative | Positive |
| 11 | Negative | Negative | 0.183 | 0.817 | Positive | Negative |
| 12 | Positive | Positive | 0.727 | 0.273 | Positive | Positive |
| 13 | Negative | Negative | 0.183 | 0.817 | Positive | Negative |
| 14 | Positive | Positive | 0.727 | 0.273 | Positive | Positive |
| 15 | Negative | Positive | 0.620 | 0.380 | Negative | Positive |

**FIGURE 11:** Screenshot of the output when the system shown in Figure 10 was implemented in RapidMiner.

Nirmalya Thakur & Chia Y. Han

|  | true Positive | true Negative | class precision |
|---|---|---|---|
| pred. Positive | 68 | 26 | 72.34% |
| pred. Negative | 28 | 79 | 73.83% |
| class recall | 70.83% | 75.24% |  |

accuracy: 73.13%

**FIGURE 12:** Performance Characteristics, represented as a confusion matrix, of the system shown in Figure 10, when implemented in RapidMiner.

The performance characteristics of this learning model as shown in Figure 12, shows that the overall performance accuracy achieved by the model is 73.13%. The respective class accuracies for predicting positive and negative user experiences are 72.34% and 73.83% respectively.

## 4. DISCUSSION OF RESULTS

This proposed framework extends a work done by Thakur et al. [18] and address its limitations in multiple ways. Firstly, this work discusses the efficacy of the proposed framework by validating it on three datasets as compared to the work done in [18] which is more of a proof-of-concept and involved analysis of the proposed methodology on a small dataset which had limited number of training samples and test data.

Secondly, unlike [18] where just one learning approach is discussed, multiple learning approaches were implemented to infer about the best performing learning model for the given datasets and their performance characteristics have been compared in detail. This comparison shows that for the given datasets, Random Forest learning approach achieves the highest performance accuracy as compared to neural networks, decision tree learning, random forest, naïve bayes, deep learning and K-NN classifier. Therefore, the use of this learning approach is justified and thus being proposed for implementation of this framework.

Thirdly, a greater number of training samples are used from multiple datasets, and proper data pre-processing, outlier detection and outlier elimination are performed, which helps the proposed predictive model for forecasting user experiences to achieve an overall accuracy of 73.13%. This is much higher than the overall performance accuracy of 67% as discussed in [18].

Finally, this framework also introduces a novel approach of analyzing the emotional response associated to different atomic activities and their context attributes based on emotion analysis from facial images. Such an approach would help to identify the emotional states after every task or sub-task related to the given activity and their respective impacts towards shaping the overall emotional response of the user as a result of performing the given complex activity. This method would give a better idea of the emotional states of the user as compared to the methodology of deducing emotional state of the user, after completing the given complex activity, based on probabilistic reasoning as discussed in [18].

## 5. CONCLUSION AND FUTURE SCOPE OF WORK

For the world to be able to sustain and support the ever-increasing population of elderly people, which has been one of the primary characteristics of this century, advanced and forward-looking development policies and sound infrastructures implemented with assistive and intelligent technologies, such as smart homes, are necessary to address their increasing needs and enhance their quality of life. The future of smart homes would involve elderly people interacting with smart systems in every aspect of their day to day living. It is essential to develop intelligent technologies that would be able to make the future of smart homes "smarter" and possess the

Nirmalya Thakur & Chia Y. Han

functionality to analyze multimodal aspects of user interactions to be able to interact with users for enhancing their user experience in a technology laden environment.

With the decreasing number of caregivers and constantly increasing burden caused to the world's economy for supporting elderly people in smart homes, it is essential to develop technology-based solutions that can adapt and assist with respect to the nature of interactions performed by them to ensure a congenial living experience. In the context of IoT-based smart home environments for supporting elderly people, Affect Aware Systems possess immense potential for improving their quality of life. The essence of improving the quality of life experienced by elderly people in smart homes, lies in the effectiveness of technology-based solutions to enhance their user experience in the context of their day to day goals. Therefore, this paper proposes a framework that can predict user experiences in an Affect Aware Smart Home environment in the context of ADLs performed by elderly people. Such a predictive model would allow forecasting of user experiences even before the user engages in an activity, which would provide scope for enhancing the user experience and thus improving the quality of life of elderly people.

The proposed framework has been tested on three datasets to uphold its relevance and efficacy. The results presented uphold the relevance of this framework for predicting user experiences in the context of ADLs in a smart home for creating a smart, adaptive and assistive environment for elderly care. As per the best knowledge of the authors, no prior work has been done in this field featuring a similar approach.

Future work would involve deploying a host of wireless and wearable sensors to establish a smart and connected IoT environment for real time implementation of this proposed framework for predicting user experience in the context of ADLs performed by elderly people in the given IoT environment.